\pdfoutput=1
\RequirePackage{ifpdf}
\documentclass{JINST}
\usepackage{amsmath}

\author{Z. Moss$^1$, J. Moon$^1$, L. Bugel$^1$,  J.M. Conrad$^1$,  K. Sachdev$^2$, M. Toups$^2$, T. Wongjirad$^1$\\
$^1$Department of Physics, Massachusetts Institute of Technology\\
$^2$Neutrino Department, Fermi National Accelerator Laboratory
}
\title{A Factor of Four Increase in Attenuation Length of Dipped Lightguides for Liquid Argon TPCs Through Improved Coating}

\abstract{
This paper describes new techniques for producing lightguides for detection of scintillation light in liquid argon time projection chambers.   These can be used in future neutrino experiments such as SBND and DUNE.     These new results build on a dipped-coating technique that was previously reported and is reviewed here.  The improvements to the approach indicate a factor of four improvement in attenuation length of the lightguides compared to past studies.   The measured attenuation lengths, which are  $>$2 m, are consistent with the bulk attenuation length of the material.   Schematics for a mechanical dipping system are provided in this paper.  This system is shown to result in coatings with $<10\%$ variations.   
}

\begin{document}
\maketitle

\section{Introduction}

Flat-panel lightguides for light detection are proposed for a number of future Liquid Argon Time Projection Chambers (LArTPCs) including SBND \cite{sbnd} and DUNE \cite{DUNE}.   Scintillation light produced in liquid argon (LAr) has a wavelength of 128 nm, too short to be detected by a vast majority of current photodetectors.  Therefore, the light must be converted into the visible to be observed.    The lightguide technology takes advantage of this requirement by embedding a wavelength shifter, tetraphenyl-butadiene (TPB), into the coating of a bar that will capture and guide the light to the end.   The bars can be assembled into a flat panel that requires substantially less space than a more traditional design based on photomultiplier tubes (PMTs), such as has been used in ICARUS \cite{ICARUS} and MicroBooNE \cite{uB}.   The end of the bar can be instrumented with silicon photomultipliers (SiPMs) that have high quantum efficiency ($\sim40$\%) for the visible light and very low dark rate at the cryogenic temperature of LAr, 87 K. 

This paper focuses on a technology for producing lightguides constructed out of clear, polished acrylic bars and covered with an acrylic-embedded TPB coating.   TPB efficiently absorbs the 128 nm scintillation light and re-emits at approximately 425 nm \cite{VicTPBpaper}. This wavelength corresponds nicely with the peak efficiency of the SiPMs. Development of these lightguides has progressed over several years, and Refs. \cite{firstguidepaper, benchmarking, jinst2015} describe the past steps.   In this paper, we present improvements in technique that provide a substantial step forward in absolute brightness and attenuation length of this family of lightguides.   In Sec. ~\ref{History}, we provide the relevant historical information needed for the discussion of the improved lightguides.  In Sec.~\ref{Improve}, we report the improved method for producing the lightguides.  In Sec.~\ref{AirResults}, we present results on tests of the new lightguides in air. 
In Sec~\ref{LArPredictions}, we use the model of the lightguides found in~\cite{jinst2015} to predict the attenuation length of the new lightguides in liquid argon.
In Sec~\ref{Future}, we discuss ongoing tests by SBND and DUNE on these new lightguides in liquid argon and provide a prediction concerning the attenuation length that will result from these tests.

\section{Relevant Background Information \label{History}}
 
Our most recent paper describing the technique for producing lightguides and benchmarking their performance was published in 2015 \cite{jinst2015}. We will therefore refer to these as the 2015 lightguides.    In this paper, we will compare our improved lightguides to the 2015 lightguides using many of the procedures and techniques developed in Ref.~\cite{jinst2015}.  Therefore, a brief review of the relevant information from that paper is required.
  
The lightguides are made of UTRAN UV-transmitting acrylic \cite{UTRAN}, with index of refraction of 1.49.  This is cut into bars of appropriate length and diamond polished on the sides and ends.   In our studies we report on bars that are 0.25"$\times$1.00"$\times$20.0".   The bulk attenuation length of a single bar has been reported as 160, 260 and 260 cm for 385, 420 and 470 nm light, respectively \cite{MufsonBaptista}.  The error on the measured bulk attenuation was not reported.
  
The bars are carefully annealed.   For the 2015 bars (and the new bars described in this paper), we use the annealing procedure described in Ref.~\cite{jinst2015}. The annealing apparatus consists
of an insulated tube that houses the acrylic bars and whose inner volume is warmed by a heat-gun inserted into one end of the tube. The temperature of the air near the output of the heat gun is 230$^\circ$F, while at the opposite end of the tube, the air temperature is measured to be 180$^\circ$F. Therefore, the air temperature throughout the tube follows a gradient from 230$^\circ$F to 180$^\circ$F with most of the tube well below 230$^\circ$F. 

\begin{figure}[t]
\centering
\includegraphics[width=0.3\textwidth]{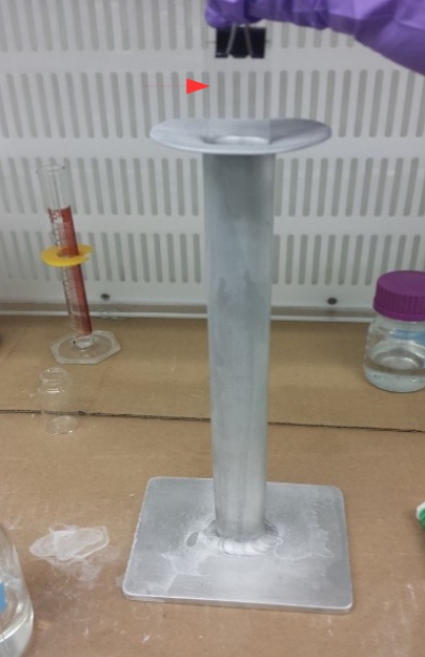}~
\includegraphics[width=0.7\textwidth]{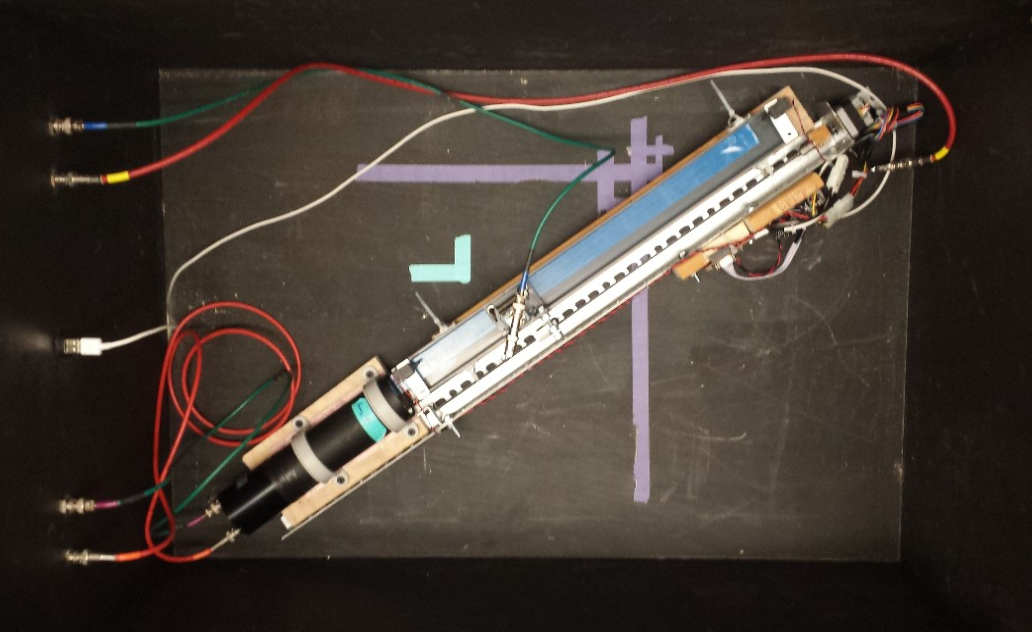}
\caption{\textit{Left: Acrylic guide being dipped in the candlestick filled with TPB coating;
Right:  Attenuation test setup. Images from Ref.~\cite{jinst2015}.}}
\label{fig:Candlestick}
\end{figure}

After the annealing procedure, the bars are thoroughly cleaned with ethanol. Next, a vertical cylinder with oval cross section large enough to contain all but the upper few centimeters of the bar is filled with the liquid coating.  This cylinder is referred to as the ``candlestick" (see Fig \ref{fig:Candlestick}, left).  The bar is then inserted into the candlestick and allowed to soak in the liquid.  Finally, the bar is drawn out vertically and allowed to air-dry.  

In Ref.~\cite{jinst2015} we had found that the lightguide performance did not correlate with the order in which up to 5 lightguides were dipped in the same batch of solution.
We now find, though, that particulates may contaminate liquid that has been used many times. Therefore, we recommend pouring the liquid out after five dips and either filtering or making a new liquid batch.  Also, to minimize particulate in the candlestick, the candlestick is now washed with ethanol immediately before the dipping process as well as after the dipping process, as was done for the 2015 lightguides.

The recipe that was used for the dipping liquid for the 2015 lightguides combined 1 g acrylic pellets, 0.5 g TPB, 50 mL toluene and 10 mL of ethanol.  The acrylic pellets assure that the coating's index of refraction matches well to that of the bar.   The TPB becomes embedded in the acrylic matrix and shifts the 128 nm light.  The toluene is the primary solvent.   The added ethanol produces a more uniform coating when the bar dries, which is essential to a long attenuation length.   The amount of ethanol required in the mix is determined experimentally by testing the attenuation length of the lightguides.

The attenuation length of the lightguides was tested using an attenuation length tester described in detail in Ref.~\cite{jinst2015} and shown in Fig \ref{fig:Candlestick}, right.   This tester uses a 286 nm LED.  The LED is moved along the bar by a stepper motor in well-calibrated steps. The LED is pulsed $10^4$ times, and the wavelength-shifted light exiting the end of the bar is measured by a PMT. The waveforms of these PMT pulses are recorded, and their integrals are histogrammed. The brightness value corresponding to a particular distance along the bar is the mean of this distribution of integrals (charges, in ADC$\times$ns). The corresponding standard error on the mean is small due to the large number of measurements and is negligible in comparison to systematic errors on the charge. The set of mean charges recorded at each distance is then fit to extract bar performance parameters. We also developed a model to describe the attenuation length measurement \cite{jinst2015, benmodel}.

The attenuation length measurement in air is sensitive to the thickness of the coating on the bar because the 286 nm light penetrates 
into the bar and excites all of the TPB in its path.  Thus, with a thicker coating, more TPB will be excited and the light output of the bar will increase. This is unlike the expected behavior of the bar when exposed to 128 nm LAr scintillation light, which does not penetrate beyond near the surface of the coating.   As a result, because the dipping process tends to yield a thicker coat at the base of the bar than at the top of the bar, the coating thickness affects the measured attenuation length. If the attenuation length is measured with top closest to the light detector and base farthest from the light detector (``forward''), the result will be artificially long due to the increase in coating thickness.    On the other hand, if one reverses the bar, so that the measurement proceeds with base nearest to the photodetector and top farthest (``backward''), one obtains an attenuation length that is artificially short.   Neither will represent the actual attenuation length as obtained when 128 nm light hits the bar.  As described in the Sec~\ref{airmodel}, one can fit the combined forward and backward measurement to simultaneously extract  parameters that describe the change in coating thickness along the bar as well as the actual attenuation length.      

The attenuation length of the propagating visible light in the lightguide depends on two properties: (1) the bulk attenuation in the lightguide and (2) surface losses.   
The first property is not related to whether the lightguide is immersed in air or liquid argon, since it is simply a property of the acrylic.   The second property leads to a change in the measured attenuation length in air versus liquid argon because the difference in indices of refraction of air and liquid argon result in a change in the angle of total internal reflection.
In Ref.~\cite{jinst2015}, we developed a simulation which uses the forward and backward measurements as inputs, and then propagates the visible light through the lightguide with a parameter to describe the fractional light loss per reflection off of the coated surface.  This paper showed excellent agreement between the model and measurements of the attenuation length of the lightguides in liquid argon.    The 2015 lightguides had attenuation lengths in the 50--60 cm range in liquid argon \cite{jinst2015}.

The bars discussed in this paper are produced of the same UTRAN acrylic, cut and annealed as described above.   The dipping process is the same as for the 2015 lightguides, with a small change to the candlestick design to reduce humidity, described below.  However, the coating recipe has changed substantially, as discussed below.    The same attenuation tester is used, with the modification that the Alazar Digitizer was replaced with a CAEN DT5740 digitizer, reducing the time required to test each 20'' bar in one direction by a factor of six, to 20 minutes.  We compared attenuation length measurements taken with the two digitizers and found that they are consistent to better than 1$\sigma$ of the measurement errors reported in Ref.~\cite{jinst2015}.   

\section{Improvements to the Previous Techniques \label{Improve}}

Our goal was to improve brightness and attenuation length.  The brightness is affected by the ratio of TPB to acrylic in the coating.  If the 128 nm light hits acrylic, it will be immediately absorbed.  Therefore, an important goal was to increase the ratio of TPB to acrylic compared to the 2015 lightguide recipe.  However, one cannot add so much TPB that the coating loses clarity and becomes white, or light will be lost when guided down the bar.  

The attenuation length is affected by the uniformity of the coating.  Thick coatings tend to produce a nonuniform, wavy surface when they dry, which can cause light to be lost as it travels along the bar.  Thus, a goal was to make a coating that is thinner than the coating used for the 2015 lightguides, which motivated increasing the amount of toluene compared to the TPB and acrylic.   We have also found that ethanol is important to producing a smooth surface.   Bars produced without ethanol have visibly rough surfaces.  In our new recipe we have honed the relative amount of ethanol to toluene through experiment, but have not introduced a major change from the ratio in the recipe used to produce the 2015 lightguides.

\subsection{New Coating Recipe and Technique \label{newinfo}}

We have adjusted the coating recipe to the following mixture:
\begin{itemize}
\item 50 mL toluene,
\item 12 ml ethanol,
\item 0.1 g acrylic, and
\item 0.1 g TPB.
\end{itemize}
Lightguides with this coating will be called the `2016' lightguides in order to distinguish them from our previous `2015' lightguides in the discussion below.

We had found that the 2015 lightguides had consistently good response if soaked in the coating solution for 5 minutes or more.   We find that for the 2016 lightguides, the bars should soak in the candlestick for 10 minutes in order to consistently produce bars with an even coating.

\subsection{Humidity Control in Laboratory}

Trial 2016 bars produced with this improved formula seemed more sensitive to the relative humidity in the lab than the 2015 bars. Specifically, the coatings of these trial bars produced during higher relative humidity summer conditions turned visibly cloudy and gave a shorter measured attenuation length
than another batch of identically prepared 2016 bars produced when the relative humidity in the lab was lower.  
The temperature in the lab is steady at 70$^\circ$F.  The ``hand-dipped" lightguide results presented here were gathered from bars made in the lab when the relative humidity was $<20\%$.    In the mechanically-dipped case, the apparatus made use of dry gas, resulting in negligible relative humidity.

\subsection{Selecting the ``Draw Time''}

We have observed that bars removed from the candlestick quickly have a longer attenuation length (fit variable $\lambda$ described below) than those drawn out of the candlestick more slowly.  Visual inspection also indicated more apparent variation in coating across the bar for those drawn more slowly.     To test this hypothesis, we removed three 20 inch bars from the candlestick in total periods of 24 s, 15 s, and 4 s.  We found attenuation lengths of 85 cm, 187 cm and 288 cm, respectively.  Based on this study, we concluded that it is best to remove the bars from the candlestick relatively quickly.   In the results presented below,  all 20 inch bars are drawn from the candlestick in 10 s or less.

\begin{figure}[h]
\centering
\includegraphics[width=0.8\textwidth]{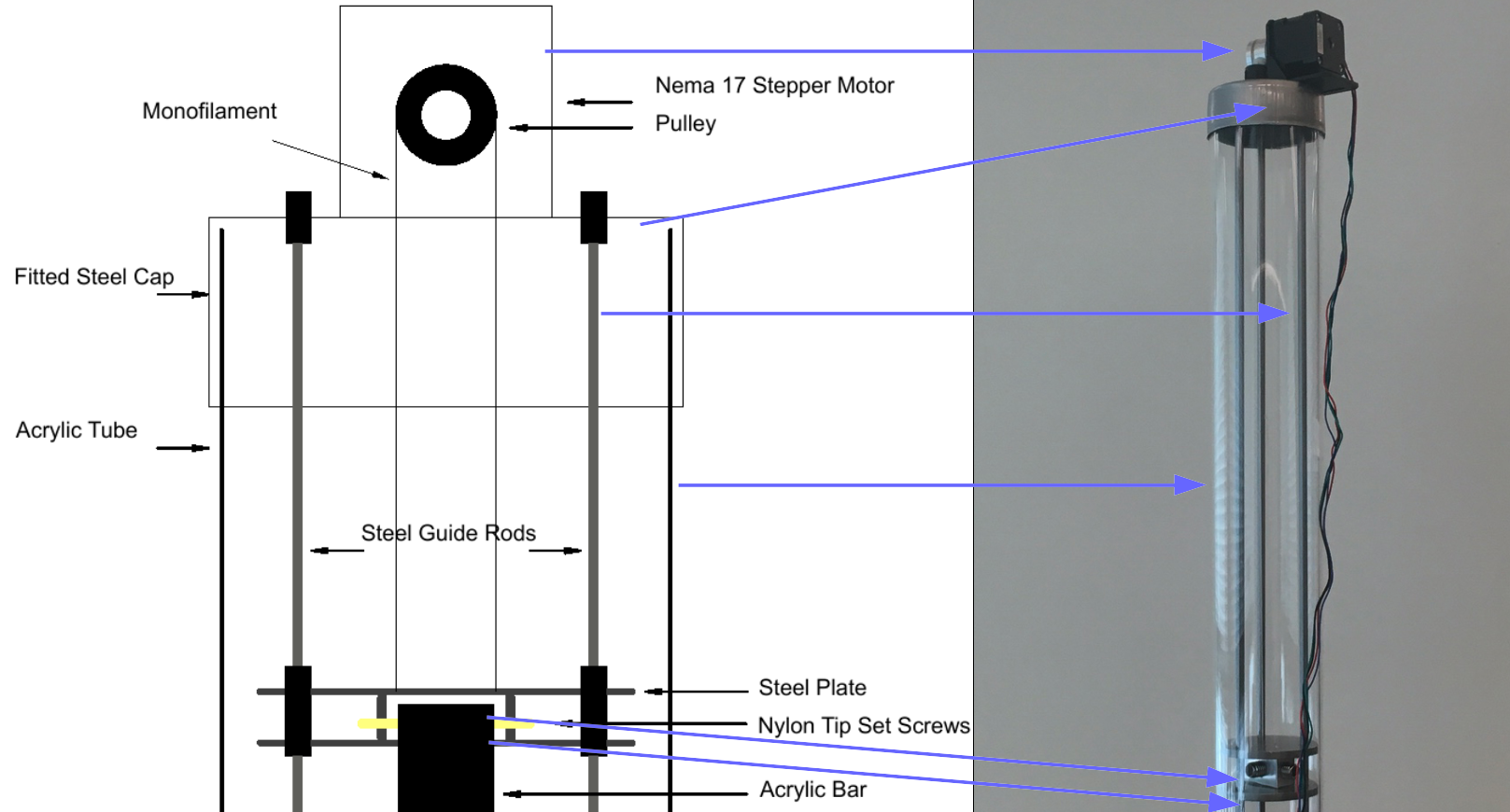}
\caption{\textit{Schematic of the drying and dipping tube }}
\label{fig:DryTube}
\end{figure}

\subsection{Mechanized Dipping System}

To further refine our production procedure, we constructed a container that automatically dips and dries the bars in near zero relative humidity conditions. See Fig.~\ref{fig:DryTube} for a schematic sketch and photograph of the dry tube setup.

The apparatus body is composed of a hollow acrylic tube oriented vertically. 
The candlestick is inserted at the base of the tube. A metal cap with a Nema 17 stepper motor attached is connected via a pulley and mono-filament system to a platform which moves up and down along three steel guide rods. The acrylic bar to be dipped is attached to this movable platform using set screws. This allows us to insert a new bar and then carefully control the ascent/descent rate of the bar.  

The draw speed for the mechanically dipped bars was set to 9 s per 20'' bar.   This draw speed was limited by the motor.   As will be discussed below, in the next generation of dipping machines, a faster motor might be desirable.

To provide a low humidity environment we introduce pressurized dry argon gas into base of the acrylic tube via a small valve. The relative humidity in the drying region is monitored by an Adafruit HTU21D-F capacitive relative humidity sensor, which provides real-time monitoring. Our monitor recorded zero relative humidity at all times during the dipping procedure. However, the technical specifications for the Adafruit monitor give an uncertainty in the relative humidity of $\pm 2\%$ for an optimized range of 5\% - 95\% relative humidity. Without more detailed information on the near zero performance of the HTU21D-F model, we can only claim that all procedures were done below 5\% relative humidity.

\begin{figure}[t]
\centering
\includegraphics[width=0.8\textwidth]{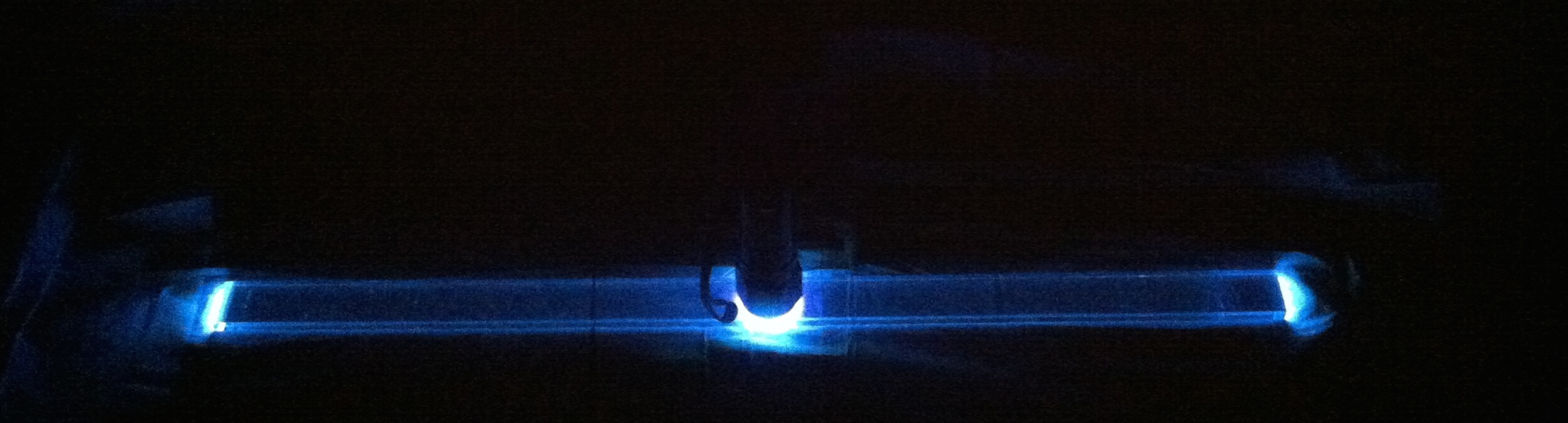}
\caption{\textit{Photograph of a 20" lightguide.  The center is illuminated by a UV flashlight.  Visible light is guided to the ends with low losses along the guide.}}
\label{fig:illum}
\end{figure}

\section{Results in Air \label{AirResults}}

\begin{figure}[t]
\includegraphics[width=1.0\textwidth]{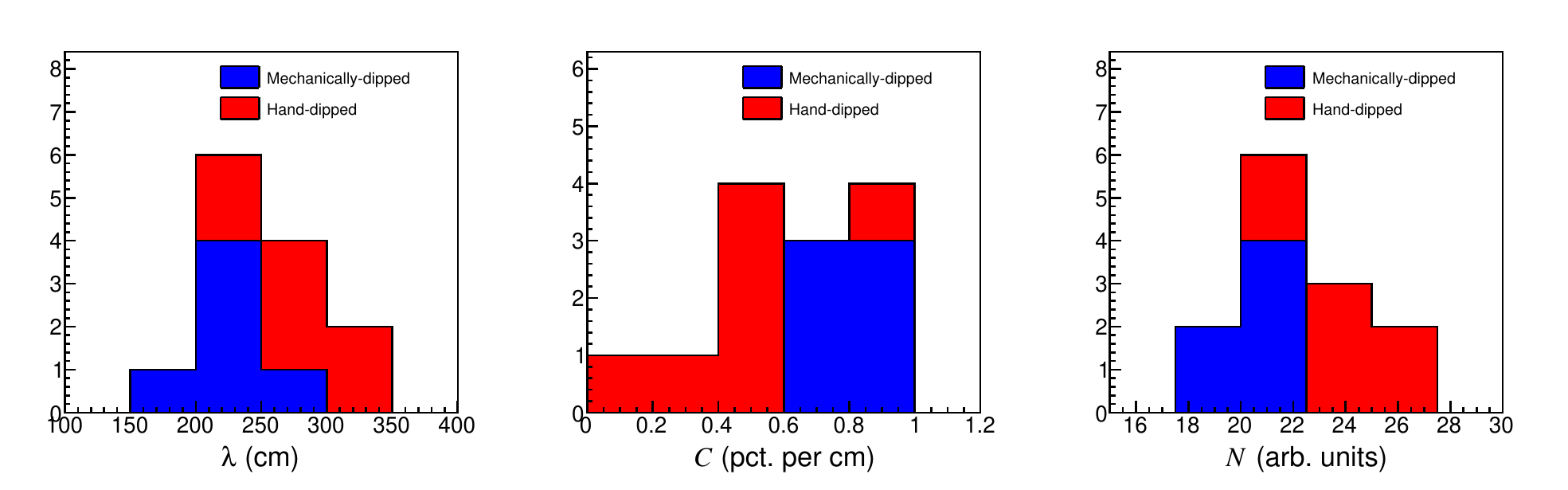}
\caption{\textit{Histograms of data presented in Tab.~1.   Red: hand-dipped lightguides,  Blue: mechanically-dipped lightguides.   Left:  $\lambda$, attenuation length;  Middle: $C$, percentage change of brightness per centimeter;  Right: $N$, normalization in arbitary units.}
\label{fig:results}}
\end{figure}

We report measurements from the two techniques:   hand dipping and mechanical dipping.   
All bars were 20'' in length.  All hand-dipped bars were produced from a single batch of coating recipe.    All mechanically-dipped bars
were produced from a different single batch of coating recipe.   Results are listed in Tab.~\ref{table:results} and shown in Fig.~\ref{fig:results}.  For each of the categories of bars, we report the average, with the standard deviation of the measurement in parenthesis.

Before discussing the results in detail, we note that these 2016 lightguides are visibly of high quality and appear substantially better than the 2015 lightguides.     Fig.~\ref{fig:illum} illustrates this point.   A UV-light is placed at the center of a lightguide.  The visible light that is captured in the bar is guided to the ends with very little loss along the bar.

\begin{table} [t]
\begin{center}
\begin{tabular}{|c|c|c|c|}
\hline
Technique & $\lambda$, Attenuation length  & $C$, Increase in thickness  & $N$,  Normalization \\ 
& (cm) & (\%/cm) & (arbitrary units) \\ \hline
\hline
Mechanically- & 229 & 0.80 & 17.6 \\
Dipped & 286 & 0.84 & 18.9 \\
& 179 & 0.77 & 20.3 \\
& 201 & 0.69  & 20.9 \\
& 213 & 0.64 & 20.7 \\
& 217 & 0.84 & 20.3 \\ \hline
mean (std. dev) & 220 (36) & 0.76 (0.08) & 19.8 (1.3) \\ \hline \hline
Hand- & 259 & 0.57 & 23.0\\
Dipped & 339 & 0.50 & 21.0 \\
 & 257 & 0.45 & 23.6 \\ 
 & 306 & 0.87 & 21.0 \\ 
 & 205 & 0.49 & 23.2 \\
 & 232 & 0.28 & 25.1 \\
 & 251 & 0.10 & 26.8 \\  \hline
mean (std. dev) & 264 (45) & 0.45 (0.25) & 23.7 (2.7) \\ \hline \hline
\end{tabular} 
\end{center}
\caption{Extracted attenuation lengths, percentage increase in brightness due to coating thickness from top to bottom of the bar, and normalization for 
two production techniques:  hand-dip (top) and mechanized-dip (middle, bottom).    
The normalization is in (ADC counts * ns) divided by 1000 -- an arbitrary unit, but one that allows for cross comparison.    
 \label{table:results}}
\end{table}

\subsection{Model Characterizing Guide Performance in Air \label{airmodel}}

In the analysis of the response of the 2016 lightguides to 286 nm LED light measured in air, we characterize lightguide performance using a model that accounts for two effects: 1) the attenuation length of the bar due to imperfections on the surface or in the bulk and 2) the varying thickness of the coating which we assume is linear from one end of the bar to the other. To constrain the latter, the light output of the bars is measured twice in our setup in two different orientations, forward and backward, as described in 
Sec.~\ref{History}.

\begin{figure}[t]
\centering
\includegraphics[width=0.75\textwidth]{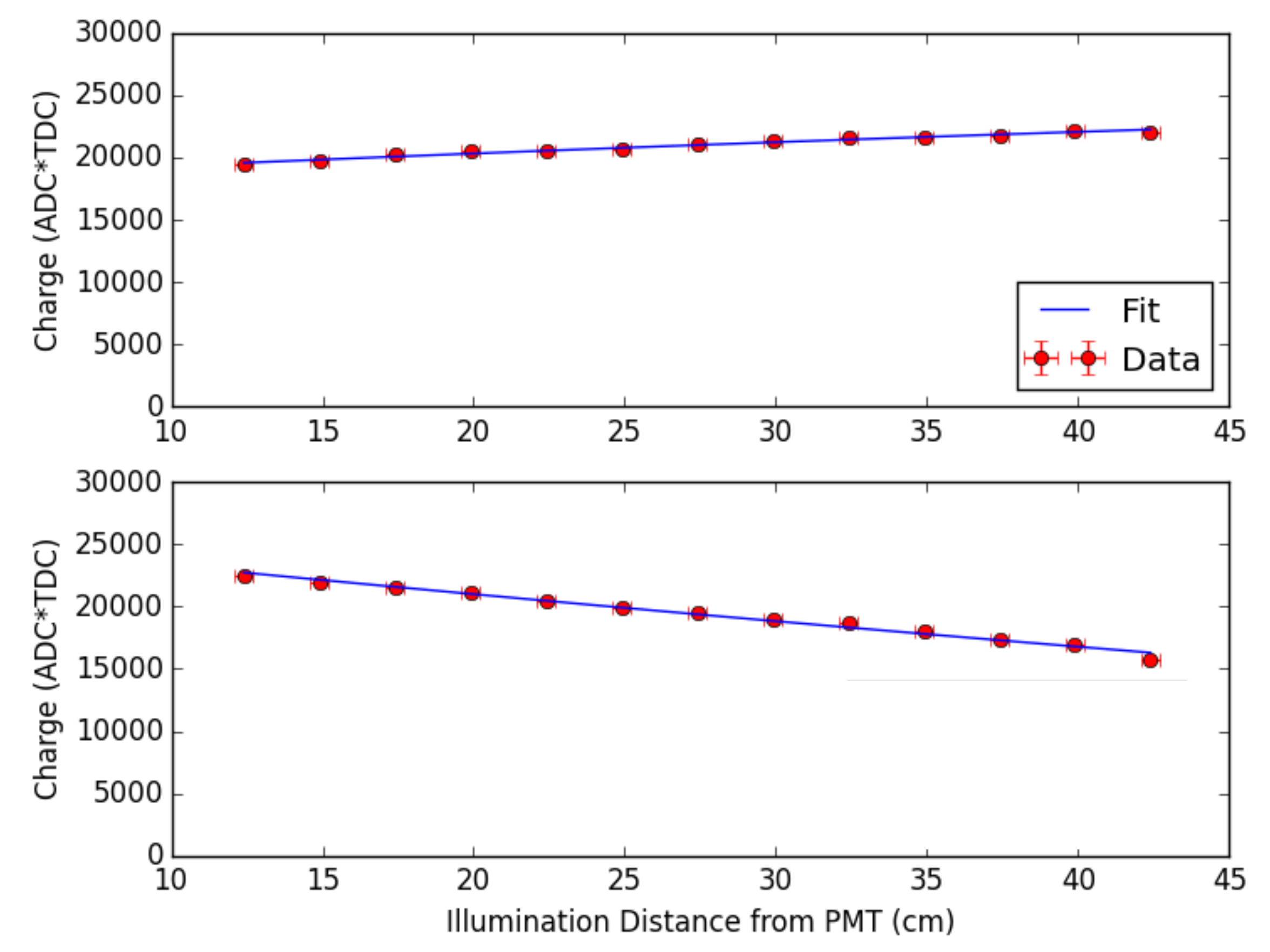}
\caption{\textit{An example of the four parameter fit to one mechanically-dipped bar. Top: forward measurement;  Bottom: backward measurement.}
\label{fig:examplefit}}
\end{figure}

We fit the forward and backward measurements simultaneously to extract four parameters.    In the forward direction, the light response is fit to:
\begin{equation}
F(x)=N exp(-\lambda*x)*(1+C*(x-m)),
\end{equation}
where $x$ is the distance from the PMT to the point measured,  $m$ is the distance from the end of the bar to the meniscus where the coating begins.   There are three free parameters in this equation: 
\begin{itemize}
\item $N$ is an overall normalization of the brightness,  measured in (ADC counts)*ns.  Because this unit is not cross calibrated to a known brightness, we call this an arbitrary unit below.
\item $\lambda$ is the attenuation length, measured in cm.
\item $C$ is the change in thickness of the coating along the bar, which is expected to increase linearly from top to bottom, and is measured in percentage change of brightness per cm.  
\end{itemize}
The backward data are fit with the equation: 
\begin{equation}
B(x)=(\epsilon*N) exp(-\lambda*x)*(1+C*(L-m-x)).
\end{equation}
where $L$ is the total length of the bar, from top to bottom.
$N$, $\lambda$ and $C$ are the three parameters simultaneously fit to the forward data.
The backward fit has an additional, fourth parameter:
\begin{itemize}
\item $\epsilon$, measured as a percentage,  allows for differences in absolute light collected by the PMT because in the backward direction the end of the bar is coated, while the in the forward direction it is uncoated. 
\end{itemize}

Of these four fit parameters, three are important to this study.
The attenuation length, $\lambda$, will be related to the smoothness of the coating and the bulk attenuation length.  This is the primary result reported about the lightguides below.  Parameters $N$ and $C$ provide measurements of the consistency of the coating between and along acrylic bars.   The results on these parameters will be shown to be more stable with a mechanical dipping system.    We find that the $\epsilon$ efficiency correction is close to unity ($>96\%$ in all cases) and not relevant to our conclusions.
When the fit is performed,  the data points are required to be within $>10$ cm and $<42$ cm to avoid the edges of the bar in the forward and backward direction.   

Fig.~\ref{fig:examplefit} shows an example fit to one mechanically dipped bar.   Note that in the forward bar arrangement, the overall slope is positive.  This is because the change in coating thickness per unit length is dominating the fit.   This is typical of all of the 2016 lightguides.

\subsection{Results on Mechanically-dipped Lightguides}

For mechanically dipped bars, we find an average (standard deviation) of the attenuation length of 
 220 (36) cm.   The 16\% spread of values encompasses the reported 260 cm bulk attenuation length 
 at 1.1$\sigma$.    Thus we conclude that the measured $\lambda$ values are consistent with the bulk attenuation length.
Through repeated measurements on the same bar,  we found that $N$, the normalization, can vary up to 2.5\%.  Systematic effects that may cause this include variations of the light source and variations in how the bar is seated against the PMT.  The PMT was allowed to ``warm up" for 30 minutes before each measurement, however variation of the dark rate after this period may also contribute to this systematic error.
The average (standard deviation) for $C$ was 0.76 (0.08) and for $N$ was 19.8 (1.3). 
Thus the spreads of both coating-specific results are $\sim 10$\% across the lightguide samples.       

\subsection{Comparison of Mechanically-dipped Results to Hand-dipped Results}

Tab.~\ref{table:results} and Fig.~\ref{fig:results} also present the results of a set of hand-dipped lightguides made with the improved 2016 coating.   
We find that the hand-dipped results are in agreement with the mechanically-dipped results within the standard-deviation for all three parameters of interest.   Interestingly, the hand-dipped results have systematically better parameter values.   However, compared to the mechanically-dipped bars, the spread in the measured values of the hand-dipped lightguides is much larger.

Consider, first, the two parameters associated with the coating, $C$ and $N$.   For $C$ we obtain an
average (standard deviation) of 0.45 (0.25) for the all hand-dipped bars.  The spread is very large -- 56\% compared to 10\% for mechanically dipped bars.    The normalization fit yields 23.7 (2.7).  This standard deviation, which is about twice as large, may be due to less reproducible behavior when hand-dipping.   Overall, the mechanical-dipping produces substantially tighter distributions of both coating-related parameters.    The mechanically-dipped and hand-dipped results agree within the spread,  but it is apparent in Fig.~\ref{fig:results}, that the hand-dipped results for $C$ are lower, and for $N$ are systematically higher, than for the mechanically-dipped case.  This indicates that the hand-dipped bars have marginally better coating performance than the mechanically-dipped bars.  This may result from differences in mixing the coating recipe, as the two samples were produced from two different batches.  Better control of the coating production process may address this spread.

On the other hand, the hand-dipped and mechanically-dipped attenuation length measurements agree reasonably well, as seen in Fig.~\ref{fig:results}.  
The average (standard deviation) of the attenuation length of the hand-dipped bars, 264 (45), agrees within less than 1$\sigma$ with that of the mechanically dipped bars.  This is also in excellent agreement with the bulk attenuation length.   As with the mechanically-dipped samples, the bulk attenuation appears to dominate.    
With this said, there may be a small systematic shift to longer attenuation length in the hand-dipped case.
We had noted earlier that the hand-dipped bars were removed with a draw of $<10$ s, but that the exact time was not carefully measured as the bars were constructed.  Faster draw-time could lead to a slightly higher attenuation length.   We recommend that in a future mechanical drawing system, the draw time, which was 9 s,  be reduced to 5 s.

\section{Predicted Performance In Liquid Argon \label{LArPredictions}}

\begin{figure}[t]
\centering
\includegraphics[width=0.8\textwidth]{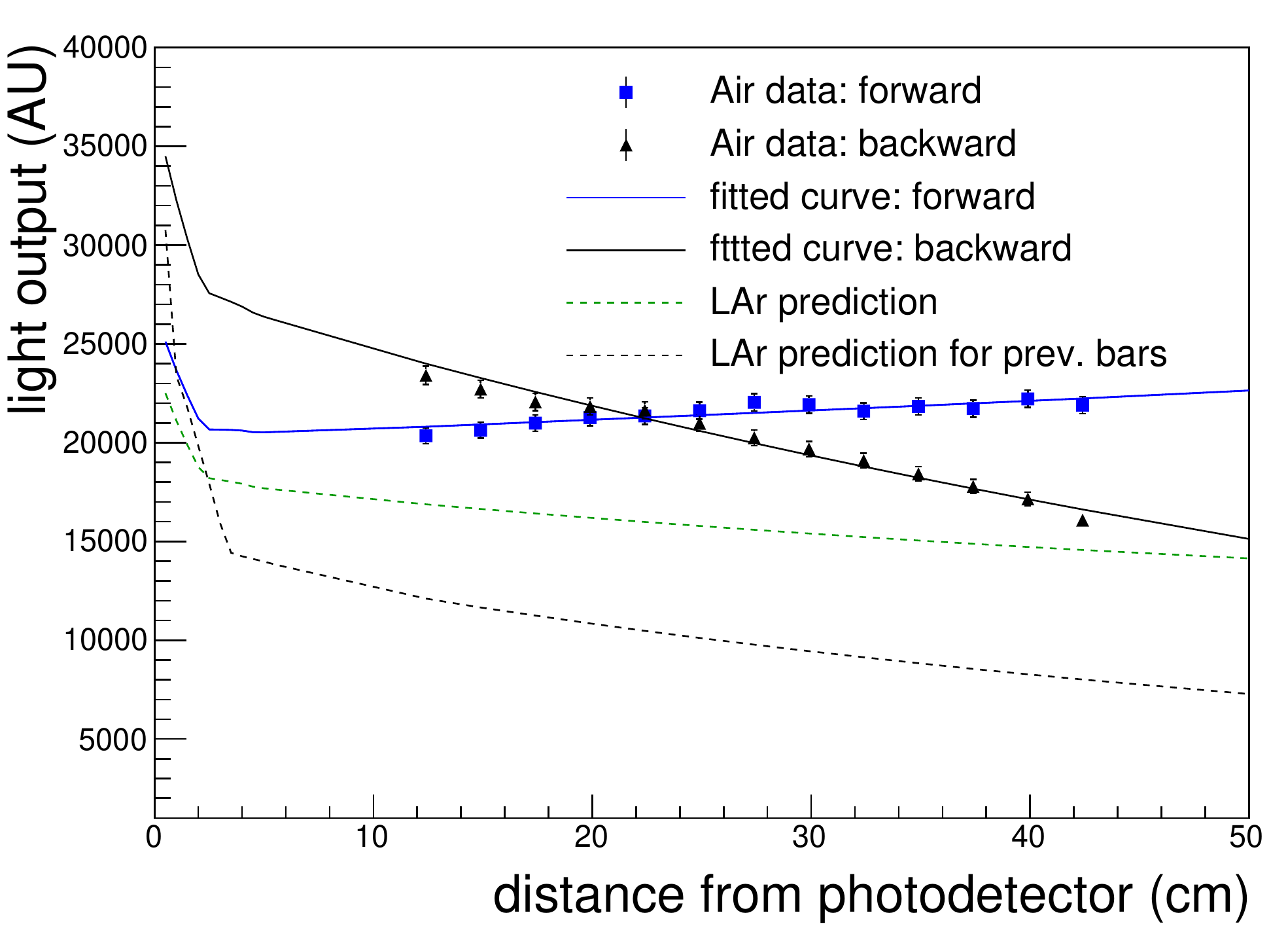}~
\caption{\textit{Predicted behavior of one of the 2016 lightguides in liquid argon. Using the simulation from Ref.~\cite{jinst2015}, we input the forward (blue solid) and backward air data (black solid) to characterize the bar and make prediction of its behavior in liquid argon (green dashed) by tracking the simulated bounces along the guide.  For comparison, the liquid argon prediction for a previous bar studied in Ref.~\cite{jinst2015} is shown as well (black dashed).} }
\label{fig:prediction_compare}
\end{figure}

\begin{figure}[t]
\centering
\includegraphics[width=0.7\textwidth]{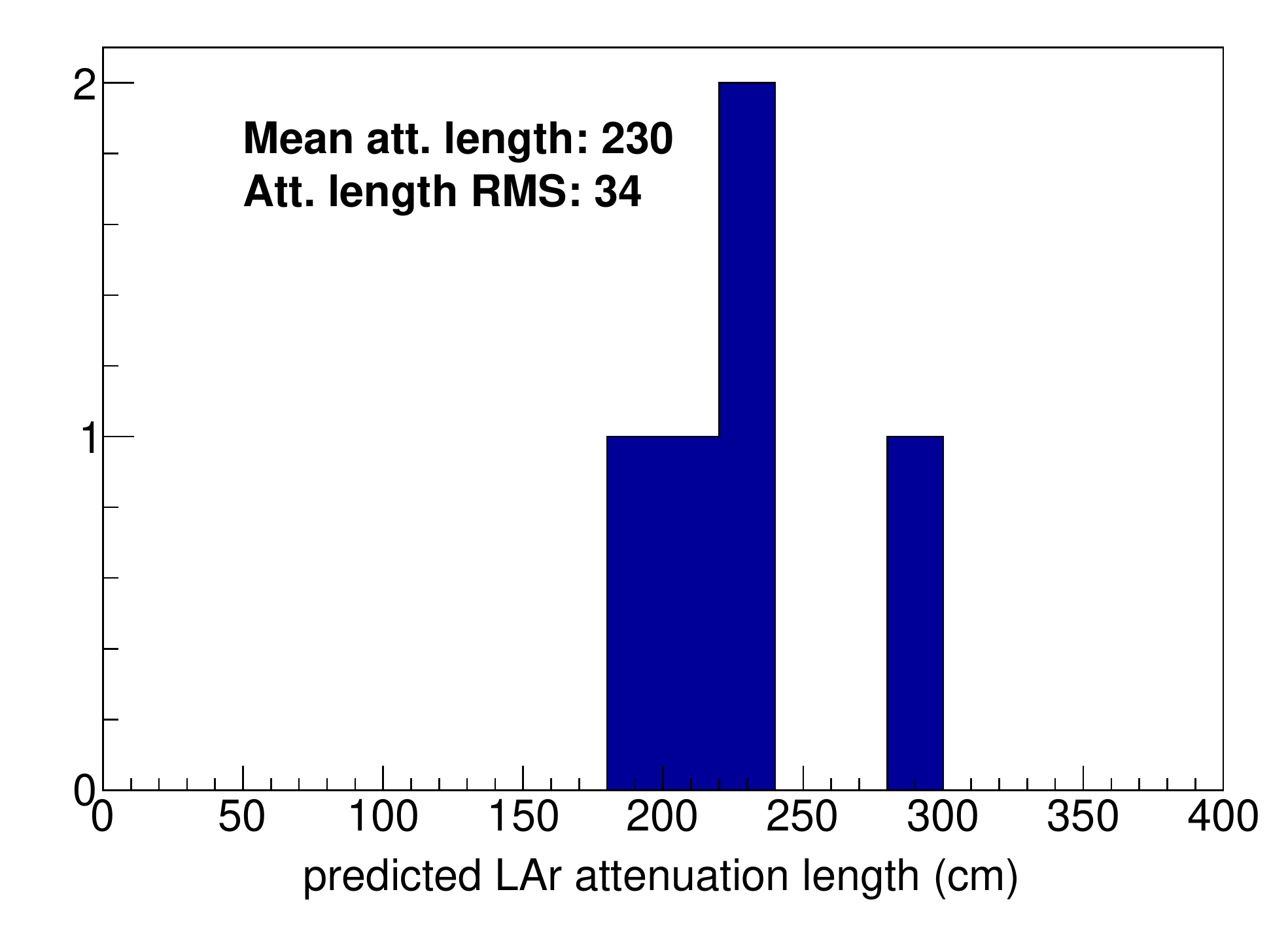}~
\caption{\textit{Predicted attenuation lengths of the 2016 lightguides in liquid argon. The simulation from Ref.~\cite{jinst2015} is used to model the bars' behavior in liquid argon using parameters found from a simultaneous fir of the forward and backward air data.} }
\label{fig:prediction}
\end{figure}

In order to verify our results, 2016 lightguides produced using the techniques of this paper are now under study in liquid argon by two subgroups, one from DUNE \cite{Stuart} and one from SBND \cite{us}.   The attenuation lengths of the lightguides will be measured using cosmic rays and sources and will improve upon the previous method by taking more granular data, thereby providing more detailed information on the attenuation length of the bars in liquid argon.  Results will be available within 6 months.    

In this section, we use the model for performance in liquid argon published in Ref.~\cite{jinst2015} to make predictions for these results.    We input the forward and backward air data to characterize the bar and make prediction of its behavior in liquid argon by tracking the simulated bounces along the guide.  
Fig.~\ref{fig:prediction_compare} shows the result of the model for the mechanically dipped lightguides in the green-dashed line.  Near the end of the lightguide, direct light hitting the sensors causes a deviation from an exponential.   But beyond about 10 cm,  an exponential can be used to quantify the predicted attenuation length of the 2016 lightguides.  As shown in Fig.~\ref{fig:prediction} we find  the bars in liquid argon are expected to have an attenuation length greater than 200 cm and consistent with the results reported in Tab.~\ref{table:results}.   The fact that the air and liquid argon results agree so well is consistent with the conclusions of the previous section that the bulk attenuation length dominates over losses from smoothness of the coating in these 2016 lightguides.     

For comparison Fig.~\ref{fig:prediction_compare} also shows the model's prediction for the liquid argon behavior of one 2015 lightguide, as the black-dashed curve.  Note that this prediction was shown to agree well with the liquid argon measurements in Ref.~\cite{jinst2015}.  One can see that there is marked improvement from the 2016 lightguides, with the green-dashed curve being significantly flatter than the black-dashed curve.  (Note that the normalizations of the green and black dashed curves are arbitrary.) The attenuation lengths of the 2015 lightguides in liquid argon were about 50-60 cm.  Thus, the 2016 lightguides have a factor of four longer attenuation length than this previous iteration.

\section{Future Step: Improved Acrylic \label{Future}}

DUNE plans to use lightguides that are 220 cm long and SBND proposes $\sim$100 cm lightguides.   These are on the scale of the bulk attenuation length of the acrylic which is now apparently dominating the light loss along the 2016 lightguides.  In this case, if the results of future tests in liquid argon bear out our prediction, then the next step to improve these dipped lightguides will be to obtain acrylic with longer bulk attenuation length than the UTRAN currently used.   Ref.~\cite{1310.6454} provides a study of bulk attenuation length of acrylic on the market, and identifies three that have bulk attenuation $>4$ m.   If this improved acrylic can be used successfully, then this will lead to the next dramatic improvement in light output of dip-coated lightguides.

\section*{Acknowledgements}

The MIT collaborators are funded by NSF grant PHY-1505855. TW gratefully
acknowledges the support provided by the Pappalardo Fellowship Program at MIT.

\end{document}